\documentclass[twocolumn]{aastex61}

\usepackage{comment}
\usepackage{amsmath}
\submitjournal{ApJ}

\shorttitle{Pollution of the solar white dwarf}
\shortauthors{Li, Mustill \& Davies}

\begin{document}
\title{Metal pollution of the solar white dwarf by solar system small bodies}

\correspondingauthor{Daohai Li}
\email{lidaohai@gmail.com,lidaohai@bnu.edu.cn}

\author[0000-0002-8683-1758]{Daohai Li}
\affiliation{Department of Astronomy\\Beijing Normal University, No.19, Xinjiekouwai St\\Haidian District, Beijing, 100875, P.R.China}
\affiliation{Lund Observatory\\Department of Astronomy and Theoretical Physics\\Lund University, Box 43, SE-221 00 Lund, Sweden}

\author{Alexander J. Mustill}
\affiliation{Lund Observatory\\Department of Astronomy and Theoretical Physics\\Lund University, Box 43, SE-221 00 Lund, Sweden}

\author{Melvyn B. Davies}
\affiliation{Lund Observatory\\Department of Astronomy and Theoretical Physics\\Lund University, Box 43, SE-221 00 Lund, Sweden}
\affiliation{Centre for Mathematical Sciences\\Lund University, Box 118, 221 00 Lund, Sweden}

\begin{abstract}
White dwarfs (WDs) often show metal lines in their spectra, indicating accretion of asteroidal material. Our Sun is to become a WD in several Gyr. Here, we examine how the solar WD accretes from the three major small body populations: the main belt asteroids (MBAs), Jovian trojan asteroids (JTAs), and trans-Neptunian objects (TNOs). Owing to the solar mass loss during the giant branch, 40\% of the JTAs are lost but the vast majority of MBAs and TNOs survive. During the WD phase, objects from all three populations are sporadically scattered onto the WD, implying ongoing accretion. For young cooling ages $\lesssim 100$ Myr, accretion of MBAs predominates; our predicted accretion rate $\sim10^6$ g/s falls short of observations by two orders of magnitude. On Gyr timescales, thanks to the consumption of the TNOs that kicks in $\gtrsim 100$ Myr, the rate oscillates around $10^6-10^7$ g/s until several Gyr and drops to $\sim10^5$ g/s at 10 Gyr. Our solar WD accretion rate from 1 Gyr and beyond agrees well with those of the extrasolar WDs. We show that for the solar WD, the accretion source region evolves in an inside-out pattern. Moreover, in a realistic small body population with individual sizes covering a wide range as WD pollutants, the accretion is dictated by the largest objects. As a consequence, the accretion rate is lower by an order of magnitude than that from a population of bodies of a uniform size and the same total mass and shows greater scatter.

\end{abstract}



\section{Introduction}\label{sec-intro}
Tens of per cent of white dwarfs (WDs) show absorption lines of heavy elements in their spectra \citep{Zuckerman2003,Zuckerman2010,Koester2014}. The stark contrast between the short sinking timescales of such material in the WD atmosphere and the long WD cooling ages suggests that accretion of metals is on-going \citep{Koester2009}. Detailed analysis showed that the accreted material is mostly dry, with compositions like that of the bulk Earth \citep[e.g.,][]{Zuckerman2007,Xu2019}. See \citet{Farihi2016,Veras2016} for reviews on observations and theories on this topic and \citet{Veras2021a} for a more recent overview.

A widely accepted scenario is that such metal accretion is caused by consumption of tidally disrupted asteroids \citep{Jura2003,Jura2008}, probably scattered onto the WD by planets in the system \citep{Bonsor2011,Debes2012,Frewen2014,Mustill2018,Smallwood2018,Smallwood2021,Veras2021,Li2021}. For instance, a planetary system, stable over the main sequence, may be destabilised during the WD phase because of the increased planet-star mass ratio and hence the increase in the interplanetary forcing compared to the central gravity \citep{Debes2002,Veras2013,Mustill2014,Veras2016a,Mustill2018,Maldonado2020,Maldonado2020a,Maldonado2021}. The planets may then interact with the small bodies in the system, sending a fraction of them to the WD \citep{Frewen2014,Mustill2018}.

In the context of the solar system, the giant planets will likely remain stable over the entire course of the main sequence, giant branch and WD phases \citep{Laskar1994,Duncan1998}. \citet{Debes2012} proposed that when the Sun becomes a WD, the mean motion resonances with Jupiter located in the main asteroid belt will widen owing to the increased planet-star mass ratio. Through numerical simulations, they found that the accretion rate is two/three orders of magnitude lower than that of the observed extrasolar WDs. \citet{Smallwood2018} discussed the shift of the secular resonance caused by the engulfment of terrestrial planets by the Sun during the giant branch and found that the resonance sweeping causes a greater extent of accretion than the mean motion resonance. Perhaps to a lesser degree of relevance to the solar system, \citet{Bonsor2011} looked at how a planet passes objects from an outer (extrasolar) trans-Neptunian belt inward and toward the central host; \citet{Smallwood2021} studied the role of two types of resonance in a broader exoplanetary background; \citep{Veras2014a} investigated how (extrasolar) Oort cloud objects can be accreted with the help of the galactic tides and stellar flybys.

The above mentioned works have mostly looked into the accretion of the solar WD from a single small body population. In this work, as expanded below, we study the accretion from three major small body populations $\lesssim100$ au in a consistent manner.

Between the orbits of Mars and Jupiter lie the main belt asteroids (MBAs). The formation of the belt, for instance how much material forms in-situ and how much has been transported and from where, is still under debate \citep[see][for a recent review]{Raymond2020b}. The dynamics of these objects is complicated as evidenced by their semimajor axis distribution where prominent dips called Kirkwood gaps are observed \citep[see][for example]{Lecar2001}. These gaps are caused by mean motion resonances and/or secular resonances \citep{Wisdom1985}, where an object's eccentricity can be pumped up such that it reaches the terrestrial planets or the Sun \citep{Wisdom1982}.

The Jovian trojan asteroids (JTAs) share the same trajectory as Jupiter but with orbital phases either leading or trailing by roughly $60^\circ$. They have been captured by Jupiter during its early orbital migration \citep{Nesvorny2013,Pirani2019}. Those objects are slowly diffusing and leaking from the trojan region and perhaps $\sim20\%$ have been lost over the age of the solar system \citep[e.g.,][]{Tsiganis2005a,DiSisto2014,Holt2020}. The escapees wander around in the solar system typically for a few Myr and a few per cent collide with the Sun \citep{DiSisto2019}.

The trans-Neptunian objects (TNOs) reside farther out. This population can be further divided into several subpopulations depending on their orbital dynamics \citep{Gladman2008}. Except for the cold classicals on low inclination orbits, the other subpopulations are transported by Neptune during its radial migration \citep[see review by][]{Morbidelli2020}. Depending on the subpopulation, these objects are gradually leaking at different rates \citep{Munoz-Gutierrez2019}, some ejected and some reaching the giant planet region. The latter may be further scattered in \citep{Levison1997}, a fraction hitting the Sun \citep{Galiazzo2019}.

The Sun will stay on the main sequence for several Gyr quiescently and will shed half of its original mass on the giant branch and become a WD \citep{Sackmann1993,Schroder2008}. The inner terrestrials will likely be engulfed by the Sun during the asymptotic giant branch \citep[the exact fate depends on the solar structure and mass evolution][]{Mustill2012}, while the four giant planets will remain stable during the entire evolution \citep{Laskar1994,Duncan1998}. As it is the giant planets that dictate the evolution of the small body populations discussed above, objects therein will probably continue to fall to the Sun during the WD phase, causing metal pollution. In this work, we aim to quantify this process with extensive numerical simulations, providing a concrete benchmark of WD accretion.

The paper is organised as follows. In Section \ref{sec-sim}, we describe the small body population used in this work and the simulation setup. Then in Section \ref{sec-res}, we analyse the result of the simulations and calculate the solar WD's accretion rate. Section \ref{sec-dis} present the discussions and conclusion.

\section{Simulations}\label{sec-sim}
In this work, we simulate the orbital evolution of small bodies during the Sun's giant branch and WD phase. To Fully track the solar system evolution during the Sun's main sequence is beyond computational capabilities. As we discuss in Section \ref{sec-dis}, the small bodies' evolution during that phase is probably mild. Similarly, the terrestrial planets are omitted to save computational time and also, these will be probably consumed during the giant branch. Finally, objects within a few hundreds of au from the Sun are immune from the perturbation of stellar flybys and galactic tides. Therefore, our model consists of the Sun+four giant planets+small bodies as detailed below.

\subsection{Small body population}\label{sec-sim-pop}
Among the solar system small body inventories, we consider the observed objects from the three major long term stable populations: the main belt asteroids (MBAs), the Jovian trojan asteroids (JTAs), and the trans-Neptunian objects (TNOs), all taken from the minor planet center \footnote{\url{https://minorplanetcenter.net}} as of 2021 March. The orbits of the giant planets are obtained from the JPL horizon system \footnote{\url{https://ssd.jpl.nasa.gov/?horizons}}. 

Hundreds of thousands of MBAs are listed in the minor planet center. Here we choose to take objects that have a semimajor axis $a$ below 4.8 au (Jovian $a$ minus its Hill radius), pericentre $q$ and apocentre $Q$ between Earth and Jupiter. This choice means that the Hungarias, Cybeles, and Hildas are effectively included (and all are loosely referred to as MBAs in this work). Also, we only consider those with an absolute magnitude $H<12.8$, corresponding to a physical radius greater than 6 km assuming an albedo of 0.1 (see Figure \ref{fig-acc-time} below). This cut is rather arbitrary but helps to limit the number of objects close to that of JTAs as we will see below. We note there is great scatter in the asteroids' albedos and densities \citep[for instance][]{Pravec2012,Carry2012}, so our selection is complete in neither size or mass. A total of 5266 MBAs are so selected in this work.

Spanning a range of heliocentric distances both today and perhaps upon formation \citep[see review by][]{Raymond2020b}, those MBAs have different physical properties as inferred from their spectral types \citep[see][for a review]{DeMeo2015}. We have made use of the taxonomy information from \citet{Bus2002a,Lazzaro2004,DeMeo2009} \citep[compiled by][]{Neese2010}, \citet{Carvano2010}, \citet{DeMeo2013,DeMeo2014}, and \citet{Popescu2018}. From those catalogues, we have acquired the Bus-DeMeo types for 3088 out of the 5266 objects. We note that the same object could be registered as different types in different works \citep[see, e.g.,][]{DeMeo2013}. But as we will see, from a statistical point of view, the exact categorisation of a particular object is not important.

The minor planet center lists about ten thousand JTAs. In this work, following \citet{DiSisto2014}, we take all that are numbered, totalling 5055.

Moving out to the TNOs, we simply include any object with $a>29.4$ au (Neptunian $a$ minus its Hill radius) that is observed at multiple oppositions. Because of their large heliocentric distances and the adverse observational bias, only 2904 objects fulfil this criterion. We have run a 10 Myr pre-simulation using the $N$-body code {\small MERCURY} \citep{Chambers1999} and only the 2844 stable ones are taken account of in the main simulations as described below. This pre-simulation has also allowed us to classify those objects following the convention of \citet{Gladman2008}. But we have used a simplistic way to identify resonant bodies: an object is recognised as resonant so long as its mean $a$ and eccentricity $e$ over the 10 Myr pre-simulation sit inside the width of a mean motion resonance \citep[MMR][]{Murray1999,Volk2020} up to order 4. Ideally, the resonators should be identified according to the behaviour of the corresponding critical angle \citep[e.g.][]{Gladman2008,Munoz-Gutierrez2019}. Nonetheless, for our statistical analysis, this crude approach suffices. We note the number of TNOs is only about half of the other two populations. So we have cloned the TNOs using the pre-simulation. The state vectors of the stable TNOs, as well as those of the planets, at 0, 5, and 10 Myr in the pre-simulation, are all taken as independent objects in our main simulation below; we call ones taken at 5 and 10 Myr as clones. Hence, we have 2844$\times$3 TNOs.

\begin{figure*}
\begin{center}
\includegraphics[width=0.9\hsize]{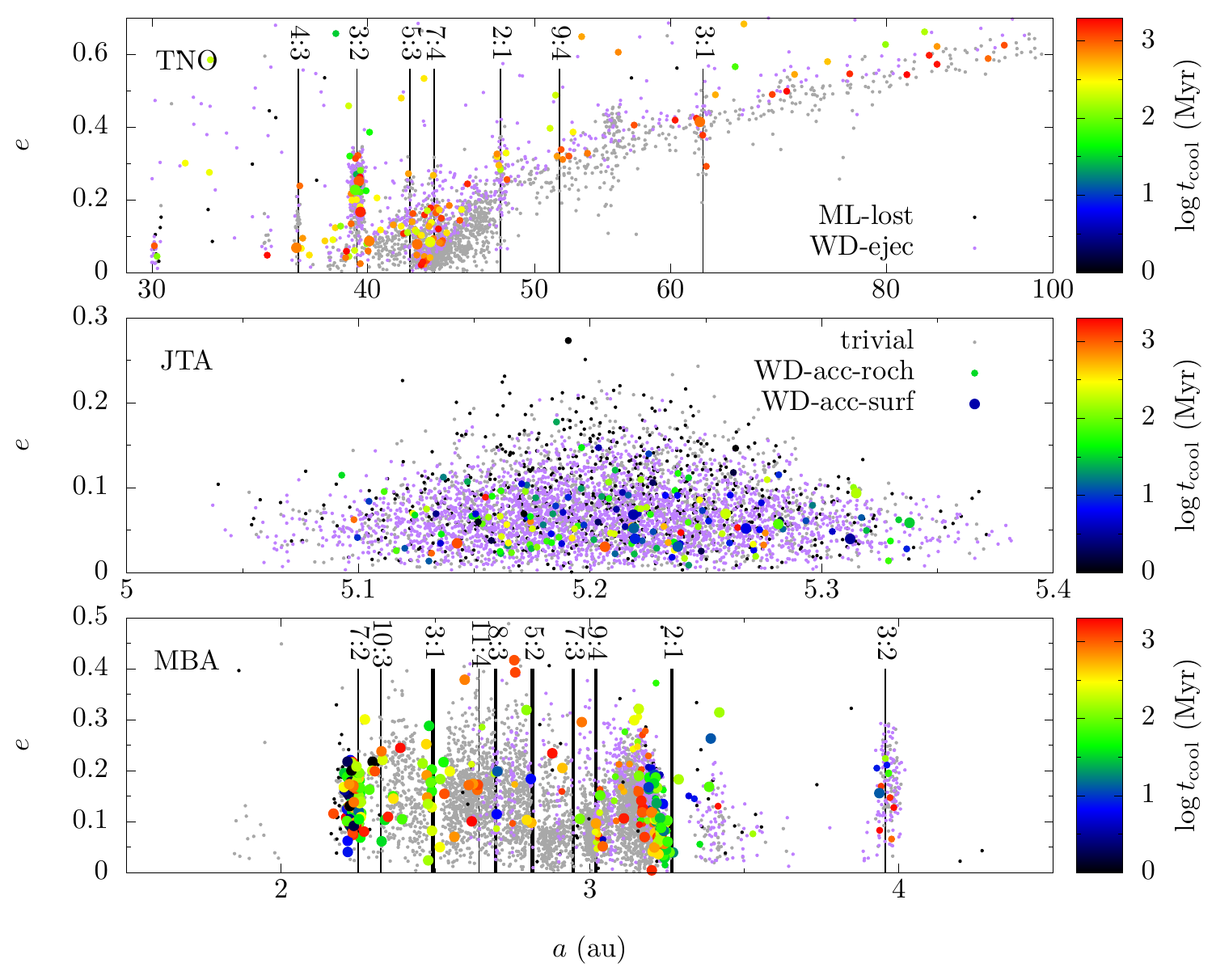}
\caption{Initial osculating orbital semimajor axis and eccentricity of the trans-Neptunian objects (TNOs), Jovian trojan asteroids (JTAs), and main belt asteroids (MBAs) studied in this work. Black dots show objects that are lost during the giant branch due to solar mass loss phase (ML-lost). Purple dots show those ejected or collided with a planet during the WD phase (WD-ejec). Coloured small and big circles represent bodies reaching the WD Roche radius (WD-acc-roch) and surface (WD-acc-surf) in the WD phase, respectively, colours meaning the timing of the event (WD cooling age) in Myr and in log scale. Otherwise, an object not involved in any of the above scenarios is plotted as a grey dot (trivial). All data are from simulations with a solar mass loss timescale of 20 Myr. For the MBAs in the bottom panel, the thick vertical black lines mark the locations of the MMRs with Jupiter associated with prominent Kickwood gaps, thin lines other MMRs where objects close-by are seen to be sent to the solar WD. For TNOs in the top panel, the vertical lines show MMRs with Neptune that seem to be able to throw bodies towards the WD; the $x$-axis is in log scale and the clones (see text for details) are not shown.}
\end{center}
\label{fig-initialorbit}
\end{figure*}

These 5266+5055+2844$\times$3 objects are shown in the $(a,e)$ plane in Figure \ref{fig-initialorbit} (TNO clones not shown). The locations of some MMRs with Jupiter in the MBA region and Neptune among the TNO region are shown as the vertical lines.

\subsection{Mass loss phase}
We omit the main sequence evolution of the solar system and start by modelling the giant branch phase. Among the other effects \citep[see][for a review; and cf. Section \ref{sec-dis} for a brief discussion]{Veras2016}, the one we examine here is solar mass loss (ML). During the asymptotic giant branch, the instantaneous stellar mass loss rate can reach a peak of $10^{-7}$ $M_\odot$/yr \citep[][; $M_\odot$ is the solar mass on main sequence]{Sackmann1993}. For planets/small bodies within thousands of au, the mass loss timescale ($T_\mathrm{ML}$) is longer than the orbital periods and their orbits adiabatically expand \citep{Duncan1998,Veras2012a}, implying that the details of the mass loss is not crucial.

Here for simplicity, we assume that the Sun's mass decreases linearly with time from $1M_\odot$ to $0.54 M_\odot$ \citep{Sackmann1993,Schroder2008} with $T_\mathrm{ML}$ being 10, 20, and 40 Myr. The change in the mass of the central host is implemented using a modified version of {\small MERCURY} by \citet{Veras2013,Mustill2018}. We have adopted the RADAU integrator and a tolerance of $10^{-11}$ \citep[see][for a discussion]{Mustill2018}. Considering the three $T_\mathrm{ML}$ values, we have 5266$\times$3 MBAs, 5055$\times$3 JTAs, and 2844$\times$3$\times$3 TNOs, providing significant statistics. In other words, we have three, three, and nine runs for the three populations, respectively.

During the ML phase, we are only interested in whether an object can remain stable within its source region and how exactly it ends is not important. So we simply remove any MBA that acquires a heliocentric distance $<1$ or $>10$ au, any JTA $<3$ au or $>200$ au, and any TNO $<3$ au or $>20000$ au.

\subsection{White dwarf phase}\label{sec-sim-wd}
After the giant branch, the solar WD slowly cools with a constant mass. We take all objects surviving the ML phase and integrate their orbital evolution during the WD phase for 2 Gyr. Along the integration, we record the close encounters between a body and the WD, when their mutual distances are smaller than the solar present-day radius ($R_\odot$) which we deem as the WD's Roche radius inside which an asteroid will be tidally disrupted \citep{Jura2003,Debes2012,Veras2014,Malamud2020,Li2021}. If the object-WD distance is smaller than an Earth radius ($R_\oplus$) which we consider as the location of the solar WD surface, the object is thought to have physically hit the WD and is accreted and removed from the simulation. However, in the former case of tidal disruption, the fragments may still be accreted later. The reason is that the planet that scatters the parent object to the WD in the first place continues to interact with the tidal fragments and throws tens of per cent of them onto the WD in a few Myr \citep{Li2021}. Therefore, we have considered both reaching the WD Roche and physical radii as criteria for accretion in our analysis in Section \ref{sec-res-acc}. Those reaching a heliocentric distance $>20000$ au are regarded ejected and is removed. A small fraction of the small bodies collide with a planet but when presenting the result, these are categorised together with the ejectees for succinctness. The above criteria for object removal apply to all three small body populations. The original {\small MERCURY} is used for the WD phase simulation and the integrator and tolerance have been Bulirsch-Stoer and $10^{-12}$, respectively.

In order to make a comparison between our derived accretion rate with the observations as done in Section \ref{sec-res-acc}, we have additionally extended the simulations for TNOs to 10 Gyr as it is those objects that dictate late accretion. But the extended simulation will only be used in that section, and the original simulations (up to 2 Gyr) are used elsewhere.

\section{Results}\label{sec-res}
\subsection{Mass loss phase}\label{sec-res-ml}
During the ML phase, the orbits of all objects adiabatically expand while eccentricity and inclination do not change \citep{Duncan1998}. During this process, while planets all remain stable \citep[][and cf. also \citealt{Veras2016b}]{Duncan1998,Veras2012a,Zink2020}, a number of small bodies are lost.

\begin{figure}
\begin{center}
\includegraphics[width=0.95\hsize]{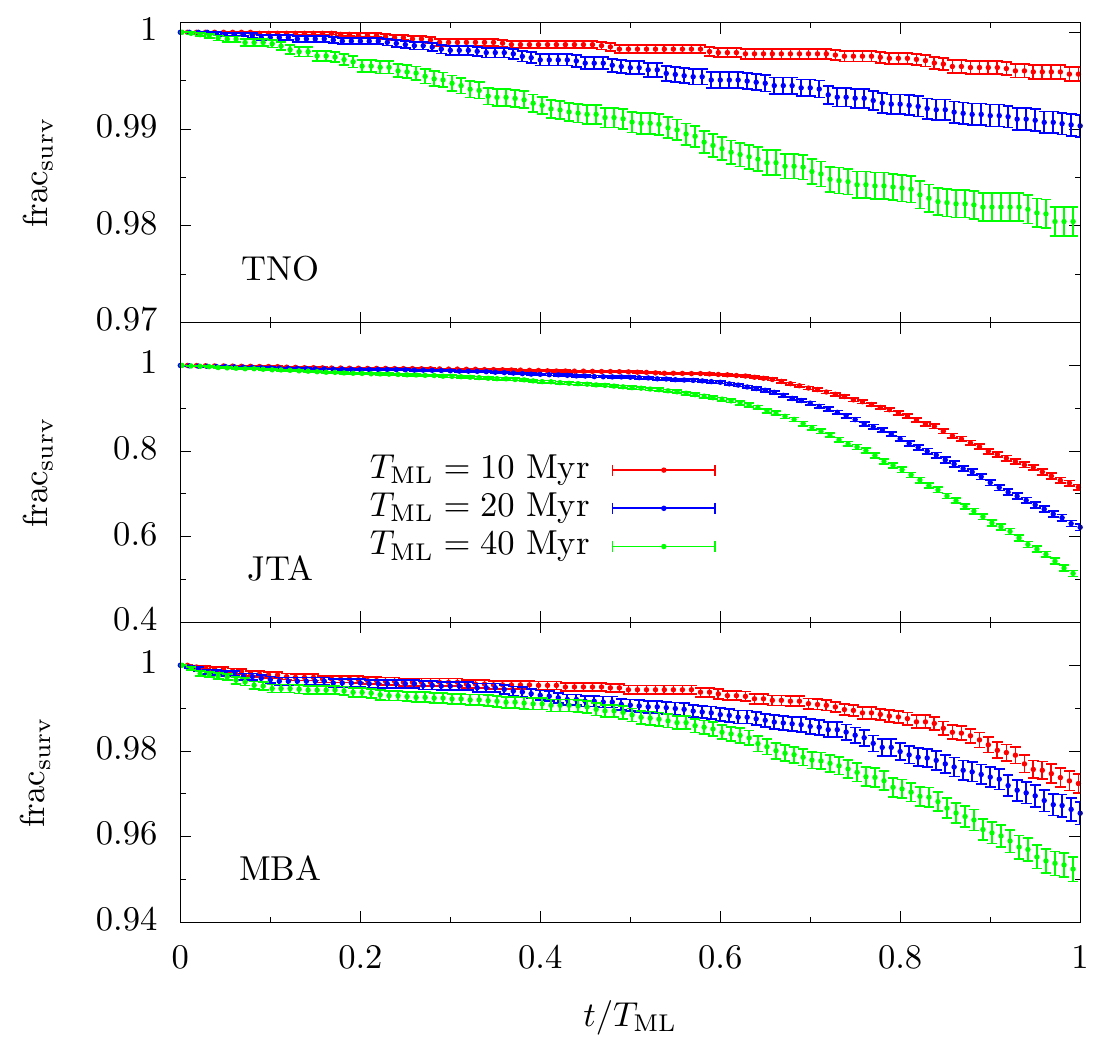}
\caption{Survival fraction of TNOs, JTAs, and MBAs during the solar mass loss phase. The fraction ($y$, number of survivor compared to the initial total number) is shown as a function of time ($x$, normalised against the mass loss timescale $T_\mathrm{ML}$, in different colours). The error bar shows $1\sigma$ dispersion as from a bootstrap method.}
\end{center}
\label{fig-gbloss}
\end{figure}

The fraction of survivors for each small body population and for different $T_\mathrm{ML}$ is shown in Figure \ref{fig-gbloss}. Overall, while the MBAs and TNOs are mostly stable with no more than a few per cent lost, 40\% of the JTAs are removed during the ML phase; and we confirm that JTAs in the leading L4 cloud are more resistant by a few per cent than the trailing L5 cloud, in agreement with \citet{DiSisto2014,Holt2020} (this phenomenon is also observed during the WD phase). Also, in the three simulations with different $T_\mathrm{ML}$ and most apparently for the JTAs, it seems that when time is measured against the respective $T_\mathrm{ML}$, the small objects tend to be lost at the same time, i.e., when the planet-star mass ratio attains some certain value.

Crossing the JTA region are numerous commensurabilities where the small bodies' orbital frequencies form small integer ratios with those of the planets \citep[e.g.,][]{Robutel2006,Erdi2007,Hou2014}. Those frequencies involve secular precession, mean motion, etc. During the ML phase, the orbits of all bodies expand in the same way, depending only on the central mass. Hence, while mean motions of all objects decrease, their ratios do not change. That is to say, MMRs do not shift. Similarly, the ratio between two secular frequencies is kept during the ML phase and the location of secular resonances will not change either. However, the strength of the resonances is enhanced because of the increased planet-star mass ratio, so objects previously lying just out of the reach of a resonance, may be affected later as the Sun is shedding mass \citep{Debes2012}. Moreover, compared to the mean motion, the secular frequency explicitly depends on the planet-star mass ratio, as it results from the interplanetary forcing \citep{Murray1999}. Therefore, during the ML phase, commensurabilities involving both types of frequencies simultaneously \citep{Robutel2006,Hou2014} may change their location, potentially sweeping the JTA region. Both the increase of the strength and the shift in location depends on the Sun's mass so an object would be affected at the same solar mass in the three simulations with different $T_\mathrm{ML}$.

Figure \ref{fig-gbloss} also suggests that a higher $T_\mathrm{ML}$ (so slower mass loss and longer simulation time) leads to a smaller survivor fraction. This can be probably linked to the higher extent of excitation associated with the slower change in frequency if resonance sweeping is at work \citep{Minton2011}; also, a longer integration allows more time for the instability to develop.

The orbitals of the objects lost in simulations with $T_\mathrm{ML}=20$ Myr are shown as black dots in Figure \ref{fig-initialorbit}. We first discuss the MBAs in the bottom panel. Apparently, most of the unstable asteroids are near the inner edge of the main belt at 2.2 au, carved by the $\nu_6$ secular resonance \citep[e.g.,][]{Knezevic1991}. On the giant branch, the terrestrial planets will be engulfed by the central host \citep{Mustill2012}, changing the secular structure of the solar system and causing the shift of the $\nu_6$ resonance and the loss of MBAs at the inner edge \citep{Smallwood2018,Smallwood2021}. Other than that, the loss of MBAs near main Kirkwood gaps \citep[for instance][]{Gladman1997} of the 2:1 MMR at 3.3 au and 7:3 MMR at 2.9 au with Jupiter can also be seen, probably caused by the increased width of the resonance \citep{Debes2012,Smallwood2021}. Sporadically, isolated asteroids from other locations are eliminated.

For JTAs, as the middle panel shows, on the contrary, there are no preferred orbits for loss or survival and objects with any $a$ and $e$ can be removed. We note we have used osculating elements and in proper element space, it is those with larger $e$ and $i$ that are preferentially lost \citep{DiSisto2014}. Then, few TNOs are removed during the ML phase (see also Figure \ref{fig-gbloss}) with obvious contribution from the Neptunian trojans and the scattered objects.

Finally, we briefly comment on the evolution of the planets. Their orbits expand like those of the small bodies \citep{Duncan1998,Veras2012a,Zink2020}. \citet{Zink2020} reported that Jupiter and Saturn will be captured into 5:2 MMR because of the increased resonance width during this process. But this is not observed in our simulations. We have also run a test simulation where the solar mass loss profile as a function of time is taken from the output of the stellar evolution code SSE \citep{Hurley2000}; no capture into the resonance is observed. It is not fully understood what has caused the difference between their and our results. We suspect if Jupiter and Saturn indeed fall in the 5:2 MMR, JTAs would be lost to a greater extent than reported here. However, as we will see, the contribution to solar WD accretion from this population is minor anyway.

\subsection{White dwarf phase}\label{sec-res-wd}
\begin{figure}
\begin{center}
\includegraphics[width=0.95\hsize]{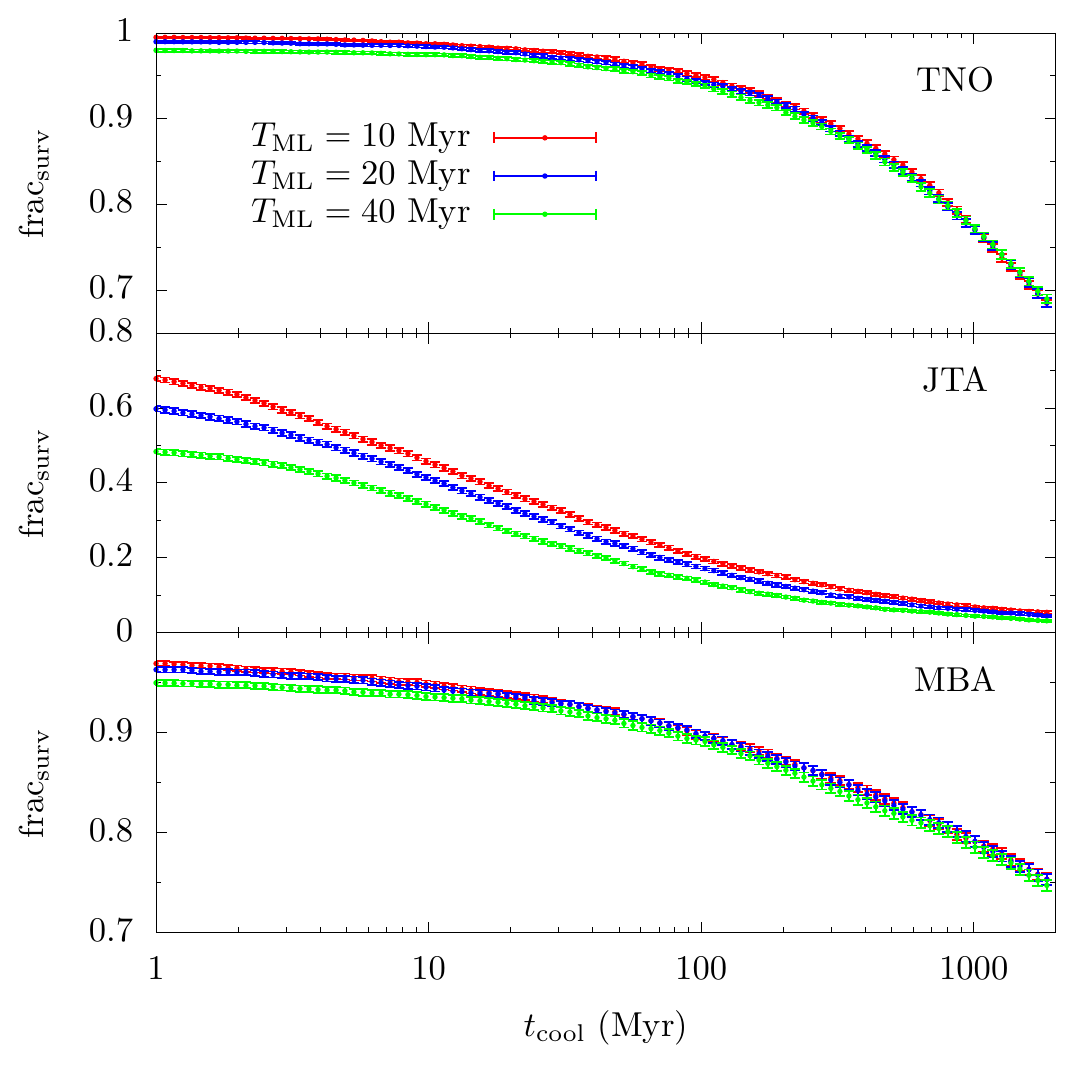}
\caption{Survival fraction of TNOs, JTAs, and MBAs during the WD phase. The fraction ($y$, number of survivor compared to the initial total number before the ML phase simulation) is shown as a function of the WD cooling age ($x$). Inherited from the ML phase, $T_\mathrm{loss}$ has been indicated in different colours.}
\end{center}
\label{fig-wdloss}
\end{figure}

The fraction of survivors (against the initial population before the ML phase) during the WD phase for the three populations and for different $T_\mathrm{ML}$ is shown as a function of time in Figure \ref{fig-wdloss}. The time ($x$-axis) here can be understood as WD cooling age $t_\mathrm{cool}$.

The loss of MBAs and TNOs is gentle and over the course of the WD phase evolution about 20\%-30\% are removed from the simulation and at end of the 2 Gyr simulation, about 70\% of the initial population still survive. In contrast, for JTAs, the removal fraction in the WD phase has been much higher $\sim50\%$ and in the end, only a few per cent remain in the simulation. Notably, the difference in the survival fraction inherited from the ML phase caused by the different $T_\mathrm{ML}$ (most apparent for the JTAs) is effectively eliminated after 100 Myr into the WD phase.

Here in the WD phase, we care about the fate of small bodies. Figure \ref{fig-initialorbit} shows the initial orbits of unstable objects for $T_\mathrm{ML}=20$ Myr. Those ejected are shown in purple dots. Those reaching the WD Roche radius (1 $R_\odot$) and surface (1 $R_\oplus$) are shown by the small and big circles, colours indicating the WD cooling age at which the event occurs.

The escape route for MBAs (bottom panel) into the inner solar system is mainly via secular resonances and MMRs \citep[e.g.,][]{Granvik2017}. MMRs not seen to destabilise during the short ML phase, for instance, the 3:1 MMR with Jupiter, can now inject a significant number of MBAs into the WD. Indeed, all MMRs associated with major Kirkwood gaps, as marked with thick vertical lines and the $\nu_6$ resonance, play a role in sending MBAs inwards. The inner MBA region (where $\nu_6$ secular resonance and 3:1 MMR are functional) witnesses a greater fraction of unstable asteroids transported toward the Sun while those in the outer region are more likely passed into the giant planet region and get ejected \citep{Gladman1997,Minton2010}. Moreover, in spite of the figure resolution, it seems that the inner instabilities tend to throw the MBAs at the Sun at an earlier time before a few hundreds of Myr though a few later than 1 Gyr are seen. The injection times of the asteroids from the outer instabilities, on the other hand, have a broader distribution. This is most obvious for the 2:1 MMR where ones closer to the resonance centre are lost early as they are within the direct reach of the MMR while those farther have to diffuse toward the MMR to be affected. Here for MBAs, and especially those from the inner instabilities, most objects that arrive at the Roche radius do reach the WD surface.

Moving to the JTAs (middle panel), it seems that they can be lost from anywhere in the $(a,e)$ plane. Those objects are slowly diffusing, transversing various resonances and increasing eccentricity and inclination, causing close encounters with Jupiter and escape from the Trojan region \citep{Robutel2006,DiSisto2019}. The escape fraction from the L5 swarm is somewhat higher than that from L4 \citep{DiSisto2014,Holt2020}. Lying in the immediate vicinity of Jupiter, the chance of collisions between the escaped JTAs with that planet becomes non-negligible, almost as frequent as that of reaching the WD Roche radius \citep{DiSisto2019}. Unlike the MBAs, only a small fraction of the JTAs that are tidally disrupted hit the WD surface (see also Figure \ref{fig-jta-acc} below). This likely has to do with the strong direct scattering with Jupiter.

The top panel shows the fate of the TNOs. The scattered objects are intrinsically unstable \citep[see][for a review]{Gomes2008} though their lifetime can be quite extended because of the resonance sticking phenomenon \citep{Duncan1997,Lykawka2007}. On leaving the scattered disk, these objects may become centaurs or Oort cloud members. Meandering around the giant planet region, the centaurs are themselves unstable on a timescale of a few Myr but could be long-surviving also with the help of resonance sticking \citep{Tiscareno2003,DiSisto2020}. While most centaurs end up ejected, several per cent reach the Sun \citep[e.g.,][]{Galiazzo2019}. The resonant objects may be chaotically diffusing, leaking from the MMR \citep{Morbidelli1997,Nesvorny2001} and they will follow the evolution of scattered objects/centaurs as discussed above. From the plot, it seems that the 3:2 MMR is especially efficient \citep[see also][]{Tiscareno2009,Munoz-Gutierrez2019}. We note that our TNO sample is nowhere near complete among those MMRs at different locations, and therefore, contributions from farther MMRs \citep[e.g., 5:2][]{Gladman2012} are not correctly accounted for because of the lack of detected objects. The classicals sitting on low-$e$ orbits roughly between the 3:2 and 2:1 MMRs with Neptune \citep{Levison1997} and also the detached objects that are not currently scattering or resonant \citep{Munoz-Gutierrez2019}, are also leaking to the inner solar system but at a rate much more slowly than, for instance, the resonant objects \citep{Munoz-Gutierrez2019}.

\begin{figure}
\begin{center}
\includegraphics[width=0.9\hsize]{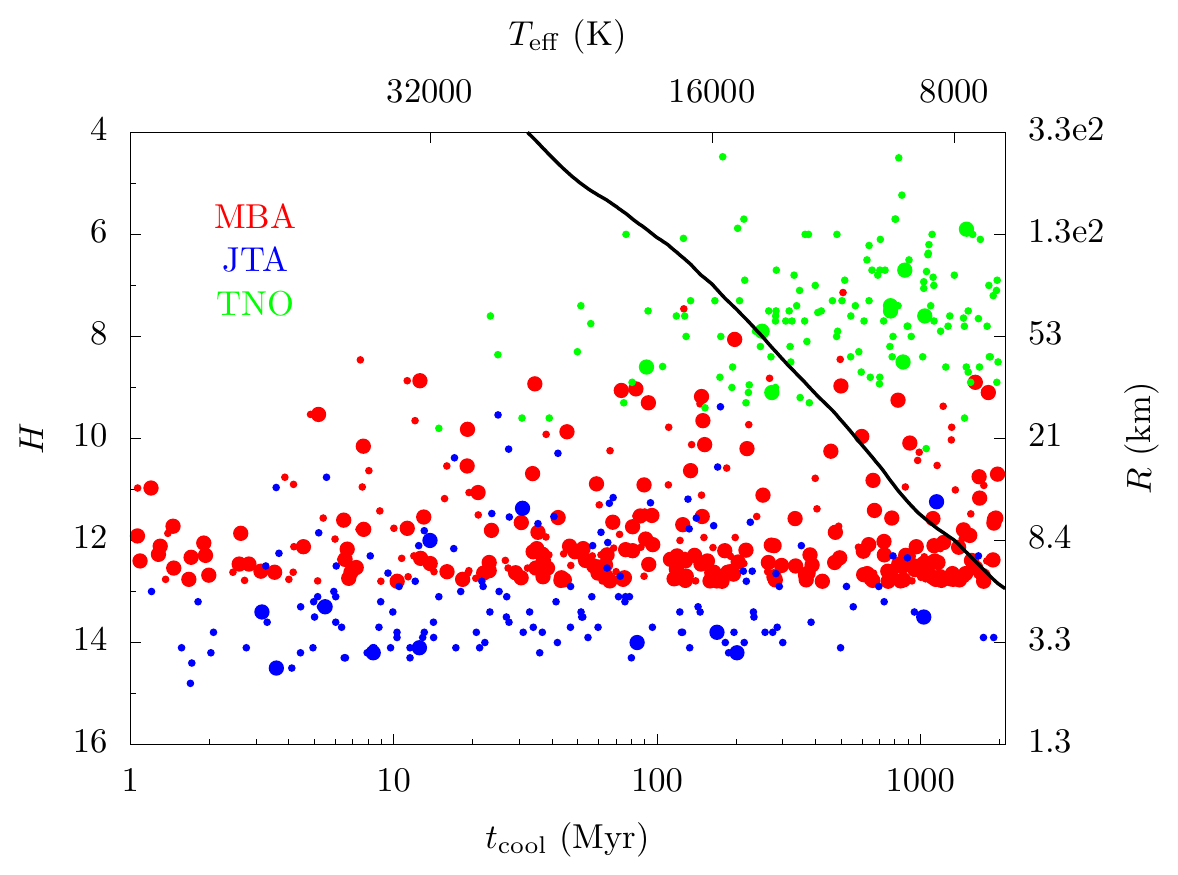}
\caption{Distribution of MBAs (red), JTAs (blue) and TNOs (green) that are accreted by the solar WD as a function of WD cooling age. The left $y$-axis shows the absolute magnitude while the right the physical radius assuming an albedo of 0.1. Small and big circles represent bodies reaching the WD Roche radius and surface, respectively. Data are taken from simulations with a mass loss timescale of 20 Myr. For the TNOs, the clones are not shown. The top horizontal axis shows the WD's effective temperature as converted from $t_\mathrm{cool}$ using equation (9) of \citet{McDonald2021} which was a fit of the model of \citet{Fontaine2001}.}
\end{center}
\label{fig-acc-time}
\end{figure}

Figure \ref{fig-acc-time} shows the distribution of the absolute magnitude $H$ (left ordinate, available from the minor planet center) of the accreted objects as a function of WD cooling age from the simulations with $T_\mathrm{ML}=20$ Myr. As the cloned TNOs are not shown, this figure essentially presents Figure \ref{fig-initialorbit} in a different dimension. The corresponding radius $R$ of each object is obtained using its $H$ and the relation \citep[e.g.,][]{Fowler1992}
\begin{equation}
\label{eq-dh}
R={664.5\,\mathrm{km}\over\sqrt{A}}10^{-0.2H},
\end{equation}
where $A$ is the albedo. Here we simply let $A=0.1$ and $R$ is marked on the right $y$-axis. As expected, accretion of MBAs (red) persists from very young WD cooling ages till Gyr old. Also, it is apparent that most of those accreted are small $R\lesssim10$ km. The hard limit of 5.8 km is imposed by the cut in $H$ that we have adopted when choosing our sample. Large asteroids of tens of km are accreted much less frequently. All accreted JTAs (blue) are small $R<15$ km and there is a clear decline as the WD ages. Consumption of TNOs begins from 100 Myr onward because of the long dynamical timescales. All TNOs arriving at the WD surface are large at least tens of km (apparent a result of our observational incompleteness) and the largest is 130 km in radius. Furthermore, a dozen of objects $\gtrsim$ 100 km (the largest $\sim$ 300 km) have entered the Roche lobe of the solar WD.

The absorbed metal has to be substantial to be observable. For instance, for a DA-type WD, a constant equivalent width of 15 mÅ for the CaII K line can be taken as the detection limit \citep{koester2006}\footnote{ The data have been extracted from their figure~1 making use of the software WebPlotDigitizer by Ankit Rohatgi ({\url https://automeris.io/WebPlotDigitizer}).} and its sharp temperature dependence implies a [Ca/H] value of $\sim10^{-6}$ at an effective temperature $T_\mathrm{eff}\sim2.5\times10^4$ K and $\sim10^{-10}$ at $T_\mathrm{eff}\sim10^4$ K. As for the WD hydrogen layer mass, we simply assume that it is $10^{-7}$ of the WD total mass, the geometric mean of the thick and thin layer models of \citet{Bedard2020}. With the hydrogen mass, the detection limit in [Ca/H] translates to an absolute calcium mass (presuming that calcium is homogeneously distributed in the atmosphere). If the calcium mass fraction relative to the small body that is accreted by the solar WD is 1\%, like observed for extrasolar WDs \citep[e.g.,][]{Xu2019}, the detection limit can be expressed in the small body mass. Then, assuming a density of 2 g/cm$^3$, the detection limit in the small body radius is acquired and plotted as a function of $T_\mathrm{eff}$ (black line) in Figure \ref{fig-acc-time}. The figure suggests that the solar WD's metal accretion may remain undetected until $t_\mathrm{cool}\sim100$ Myr (or $T_\mathrm{eff}\sim2\times10^4$ K) when 100~km TNOs are consumed. At $t_\mathrm{cool}\gtrsim$ 1 Gyr ($T_\mathrm{eff}\lesssim\times10^4$ K), accretion of small 10~km size range MBAs also become observable. However, we note that the above result, though qualitatively sensible, lacks details and should be taken with caution. For instance, the small body Calcium fraction and the WD hydrogen layer mass fraction have been taken rather simplistically \citep[e.g.,][]{Wachlin2021}.

\subsection{The solar WD accretion rate}\label{sec-res-acc}
Having established from where and when an object is accreted onto the WD, we now proceed to estimate the accretion rate for the MBAs, JTAs, and TNOs, respectively. Having set up our simulation in the real solar system, we need to determine the population's total mass.
\subsubsection{Accretion rate for MBAs}

The current total mass of the MBAs is well constrained to be $4.5\times10^{-4} M_\oplus$ \citep[$M_\oplus$ is an Earth mass][]{Krasinsky2002,DeMeo2013}. This work uses the above value as the total mass of our MBA sample which also includes the Hildas. In our first scheme, we simply divide the total mass evenly into all MBAs -- every object has the same mass as anyone else. Therefore, the number of objects accreted directly represents the mass accreted. Then we divide the 2 Gyr WD phase simulation into 25 segments evenly distributed in time in log scale and assume a constant accretion rate within each segment which is simply the mass accreted therein divided by the length of the time interval. We call this accretion rate ``count-based''.

However, we note that the mass distribution of the MBAs is highly top heavy with a few tens of objects comprising more than half of the population's total mass. In order to account for this feature, we have also adopted a second approach to calculate the accretion rate where for each body we calculate its $R$ using Equation \eqref{eq-dh} assuming a constant $A$ for all the MBAs. Then the volume of each object and hence the total volume of the entire population are obtained. Now we repeat the procedure as done above for the count-based rate but now we distribute the total population mass to each object in proportion to its volume. The accretion rate so derived is called ``volume-based''.

In actuality, the density and albedo vary significantly among the MBAs. For instance, S-type asteroids have an average albedo four times and an average density twice those of the C-types \citep{Carry2012,DeMeo2013}. We would like to take this into account and establish a perhaps more accurate accretion rate. Here, for each taxonomical type, we take the average albedo and density from \citep{Carry2012,DeMeo2013,Usui2013}. For a type without such information or an object without a taxonomical classification, we simply let its $A$ be 0.1 and density be 2 g/cm$^{3}$. Using Equation \ref{eq-dh}, we obtain an absolute mass estimate for each individual in our MBA population. The total mass thus derived is $3.5\times10^{-4}M_\oplus$, smaller than the actual mass $4.5\times10^{-4} M_\oplus$ by 30\%. We therefore multiply each mass by a common factor of 1.3 such that the total mass of our sample is $4.5\times10^{-4} M_\oplus$. Then we proceed as above to calculate the accretion rate. We note here we are not trying to derive accurate masses for every MBA. The scatter in density and albedo within each taxonomical type is significant; also, the largest objects have not been treated specifically. Here the aim is to estimate the accretion rate in a different way using more parameters than the apparently-simplistic approaches above. We call this accretion rate ``mass-based''.

As described in Section \ref{sec-sim-wd}, in the simulations, two types of ``accretion'' events have been recorded, one being an object reaching the Roche radius (Roche accretion) and the other the WD surface (surface accretion). Accretion rates based on both criteria have been calculated. For each of the MBA runs with different $T_\mathrm{ML}$, we calculate its respective accretion rates. With the caution that scatter can be large, no qualitative difference is seen between the runs so we discuss the mean rates in the following.

\begin{figure}
\begin{center}
\includegraphics[width=0.9\hsize]{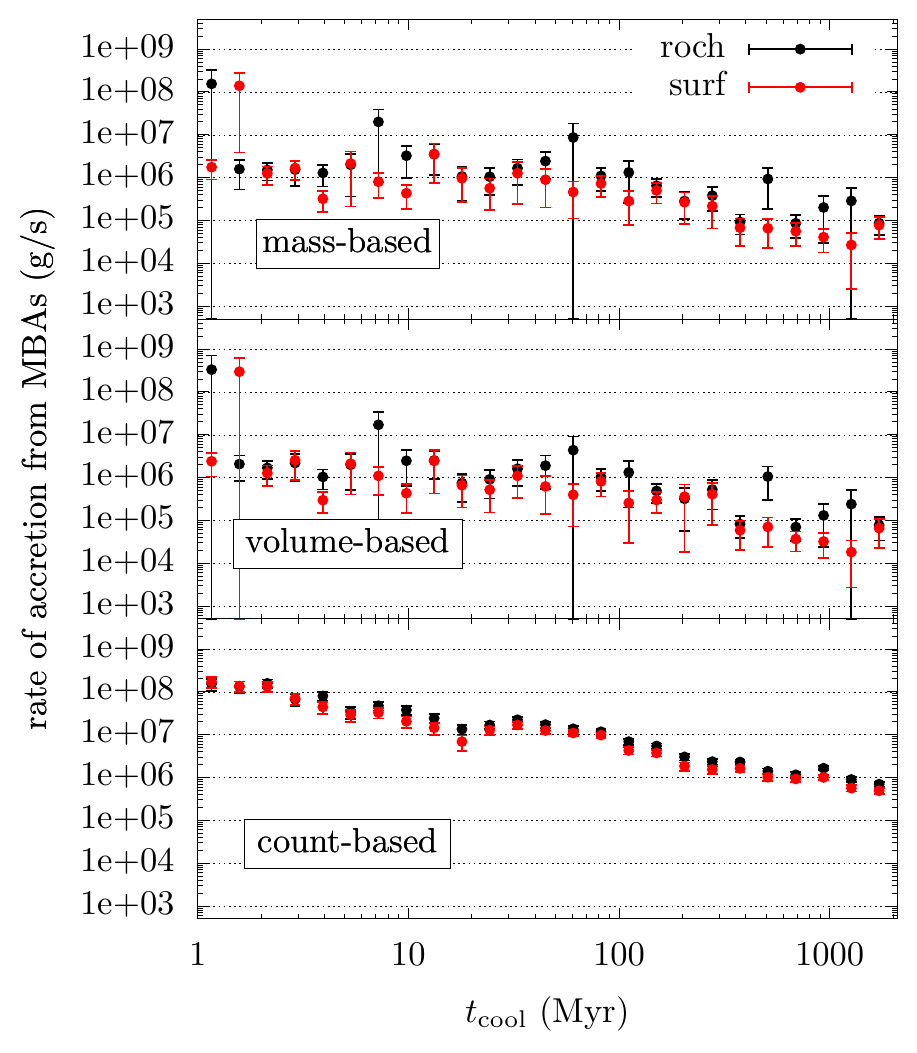}
\caption{Accretion rate of the solar WD from the MBAs as a function of WD cooling age. The black points show the amount of mass reaching the WD's Roche lobe whereas the red ones the surface. In the bottom, the rate is calculated based on the count of asteroid reaching the WD; in the middle, the rate has been corrected for the asteroid volume, assuming the same albedo; in the top, a mass-based estimate using the albedo and density by the asteroid taxonomy is adopted.}
\end{center}
\label{fig-mba-acc}
\end{figure}

All the accretion rates for the MBAs are presented in Figure \ref{fig-mba-acc}. The count-based rate, as shown in the bottom panel, is declining from $\sim10^{8}$ g/s at a cooling age of a few Myr to $\sim10^{6}$ g/s at 1 Gyr. The rates of Roche accretion (black point) and of surface accretion (red point) are within a factor of a few. The volume- and mass-based rates are shown in the middle and top panel. The two are visually very similar, which is not unexpected as in estimating the mass of an asteroid, the variation in albedo and density is at most a factor of a few and it is the volume that dominates. This volume information is captured by both methods.

We discuss the volume- and mass-based accretion rates together. Compared to the count-based rate, these two, while declining with time as well, show much greater scatter between adjacent measurements; also, these two rates are, roughly speaking, lower by almost an order of magnitude \citep[cf.][]{Wyatt2014}. Both differences stem from the fact that the MBAs follow a top-heavy size frequency distribution. Most of the time, accretion comes from the more numerous smaller less-massive objects. Hence the actual accretion rates are much lower than if all asteroids are assigned the same mass as done in the count-based approach. Sporadically, a larger more massive object is absorbed, causing local spikes in the accretion rate. Particularly, at 1.5 Myr, the surface accretion rate is higher than the Roche rate by two orders of magnitude, which is counter-intuitive. The reason is an object is registered as reaching the Roche radius the first time it does so but it remains in the simulation and it is removed until it reaches the WD surface or is ejected. Therefore, the two types of accretion event may not occur at the same time.

\citet{Debes2012} calculated the accretion rate of the solar WD from the large MBAs using a population about an order of magnitude smaller than ours. And they adopted much like a mass-based approach, assuming a uniform density. Their integration time was mostly 100-200 Myr with a few tests reaching 1 Gyr. At a cooling age of 100 Myr, our mass-based rate agrees well with theirs but at 1 Gyr ours is higher than their (much extrapolated) value by a factor of several.

\begin{figure}
\begin{center}
\includegraphics[width=0.95\hsize]{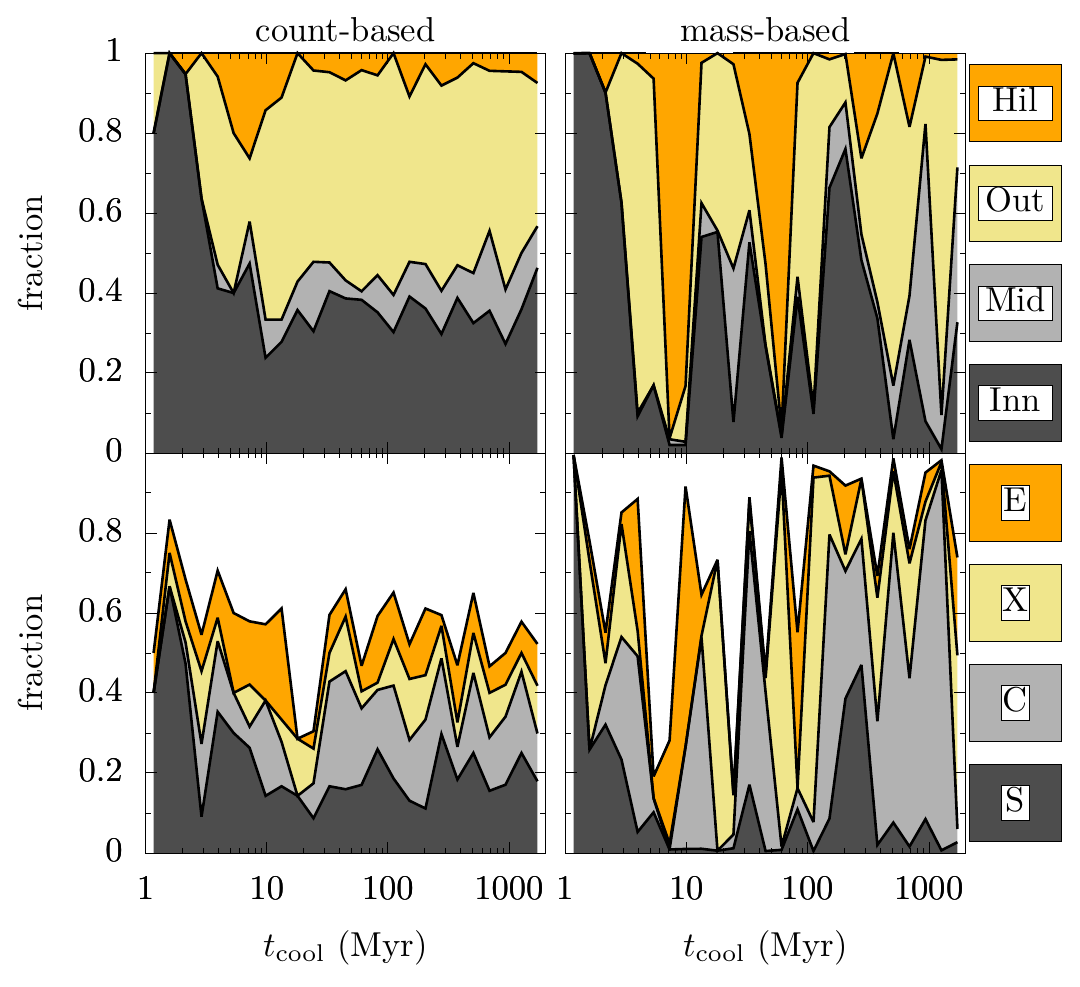}
\caption{Initial heliocentric location and taxonomy of MBAs accreted onto the WD as a function of WD cooling age. The top row shows the fraction of objects accreted from the inner (dark-grey), middle (light-grey), outer belt (khaki), and from the Hildas (orange). The bottom row shows the spectral types of the accreted MBAs: S- (dark-grey), C- (light-grey), and X-complexes (khaki), and the end members (E, orange); otherwise no taxonomical information is available (white). The left column shows the count-based accretion while the right mass-based.}
\end{center}
\label{fig-mbatypeacc}
\end{figure}

In Section \ref{sec-res-wd}, we have shown that the different mechanisms functional at different locations behind the delivery of MBAs onto the WD may operate on different timescales. Does this imply that the material accreted onto the WD would evolve with the WD cooling age as there is a gradient in the MBA composition as a function of the heliocentric distance \citep[e.g.,][]{DeMeo2014}? In order to answer this question, we take the asteroids that reach the WD Roche radius from all the simulations and classify these broadly according to their taxonomical types \citep[S-, C-, X-complex, End members and those unknown and cf. for instance][]{DeMeo2015} and to their initial locations [inner $a<2.5$ au including Hungarias, middle $a\in(2.5.2.8)$ au, outer belt $a\in(2.8.3.3)$ au, and Hildas $a>3.3$ au also including Cybeles]. The fractional contribution from these classes compared to the total accretion from all MBAs is shown in Figure \ref{fig-mbatypeacc} as a function of WD cooling age. The top row shows that for the locations and the bottom the taxonomical types. The count-based approach in the left column clearly show that within a few Myr, accretion mainly comes from S-complex asteroids and from the inner belt and after 10 Myr, C-complexes from the outer belt catch up and contribute comparably since then. This agrees with Figure \ref{fig-initialorbit} which is essentially also count-based. However, all these patterns are erased once the mass (or volume, as shown above) of the asteroid is taken into consideration as presented in the right column. Now, location-wise, the different parts of the belt are competing extensively -- any can prevail for a short while but not for an extended period of time. Taxonomy-wise, similarly, no apparent persistent features can be extracted with the possible exception of the late dominance of the C-complexes beyond a few hundreds of Myr. But as we will see, these characteristics will be anyway eclipsed by the accretion from TNOs.

\subsubsection{Accretion rate for JTAs}
\begin{figure}
\begin{center}
\includegraphics[width=0.9\hsize]{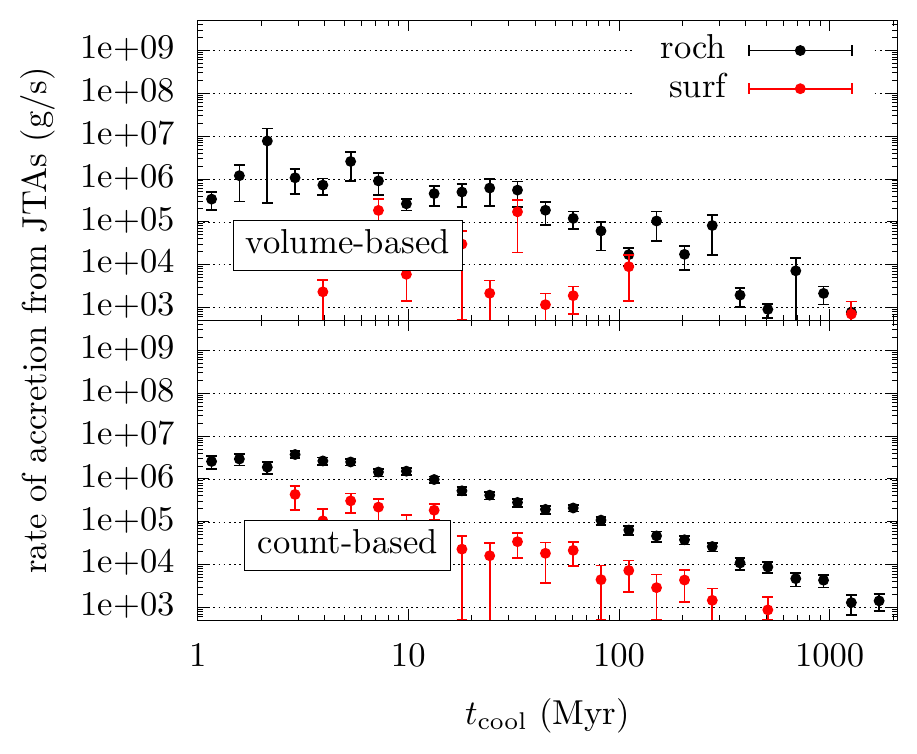}
\caption{Accretion rate of the solar WD from the JTAs as a function of WD cooling age. The symbol meaning and panel annotation are the same as Figure \ref{fig-mba-acc}.}
\end{center}
\label{fig-jta-acc}
\end{figure}

Now we proceed to calculate the accretion rate for the JTAs. The total mass of the JTAs is not so stringently constrained but seems to converge within the range $10^{-6}-10^{-5}M_\oplus$ \citep[see][for a list of historic estimates]{Pitjeva2019}. Here we simply take $10^{-5}M_\oplus$. For the JTAs, the taxonomical information is only available for a few hundreds of objects \citep[see discussion in][]{Holt2021}, not allowing for a mass-based accretion rate estimate. Also, as we have shown above for the MBAs, volume- and mass-based rates agree with each other fairly well in the main features of the accretion. Thus here we only calculate the count- and volume-based accretion rates for the three runs and the mean rates are shown in Figure \ref{fig-jta-acc}. The count-based Roche accretion rate (black points in the bottom panel) is declining consistently from a few times $10^6$ g/s for a cooling age $<$ 10 Myr to only $\sim10^3$ g/s beyond 1 Gyr. That for surface accretion is smaller by an order of magnitude. The volume-based rates show much greater scatter. Most importantly, the accretion from the JTAs is always less efficient than that from the MBAs by an order of magnitude.

\subsubsection{Accretion rate for TNOs}\label{sec-acc-tno}
Now we turn our attention to the TNOs. Here we have used our extended simulation reaching 10 Gyr. Not shown in Figure \ref{fig-wdloss}, 45\% of the TNOs still remain in the simulation at 10 Gyr. Their total mass is $\sim0.01-0.1M_\oplus$ \citep[see][for a compilation of historic estimates]{Pitjeva2018a} and we take the value $0.02M_\oplus$ \citep{Pitjeva2018a}. We note that the above quoted mass is a dynamical mass measuring the collective gravitational effect of TNOs on other bodies in the solar system and the TNOs are modelled as the largest members plus a belt representing the numerous small objects. However, our TNOs are taken from the incomplete observational sample covering a much larger space. But for the purpose of obtaining a rough estimate of the total mass of the TNOs, this simplistic treatment suffices as anyway, the difference between different estimates is within an order of magnitude. We note that we have been conservative in choosing the mass \citep[for instance, another estimate of dynamical mass by][is higher by a factor of three than that adopted here]{DiRuscio2020}.

As before, we first perform a count-based estimate of the accretion rate, assuming that the population's total mass is evenly distributed among each body. Next, the volume of each object is calculated from its diameter via Equation \eqref{eq-dh} assuming a uniform albedo. Then a volume-based accretion rate where the total mass is distributed according each body's volume is acquired.

Our TNO sample has been categorised into different dynamical subpopulations as in Section \ref{sec-sim-pop} according to \citet{Gladman2008} and we would like to use this information to obtain a somewhat more accurate accretion rate. Here we only take four subpopulations: the hot classicals, the scattered, detached, and resonant objects in that they may be characterised better and that their formation from the same origin is understood reasonably well \citep[e.g.,][]{Morbidelli2020}. The cold clsssicals are removed from our sample as they have a different origin, size frequency distribution, and a small total mass of $\sim10^{-4} M_\oplus$ \citep{Fraser2014}. Therefore, if ignoring the possibly different level of collisional evolution \citep[e.g.,][]{Abedin2021}, the four subpopulations have the same size frequency distribution. Hence, the number of objects within each subpopulation with $H$ brighter than some threshold \citep{Adams2014,Abedin2021} can be used as a proxy to assess total mass of that subpopulation relative to another. Here, we simply assume that the total masses of the scattered, detached, and resonant objects are twice, twice, and the same as that of the hot classicals, respectively, in rough agreement with \citet{Abedin2021}. Here in order to make the comparison with the above count- and volume-based rates straightforward, we just let the total mass of the four subpopulations be $0.02M_\oplus$. Then, for example, the resultant mass of the hot classicals is $0.0033M_\oplus$, less than but within a factor of a few compared to the dedicated estimate \citep{Fraser2014}. Now, each object within each of the four subpopulations (each containing at least hundreds of objects) is assigned a volume and as above, the volume-based accretion rate for that subpopulation is obtained. We perform this calculation in the hope that the samples within each subpopulation may be more homogeneous. Apparently though, for instance, the resonant objects in different MMRs suffer from different levels of observational incompleteness \citep[e.g.,][]{Gladman2012} because of their different heliocentric distances. We call the sum of the accretion rates of the four subpopulations ``revised volume-based''.

\begin{figure}
\begin{center}
\includegraphics[width=0.9\hsize]{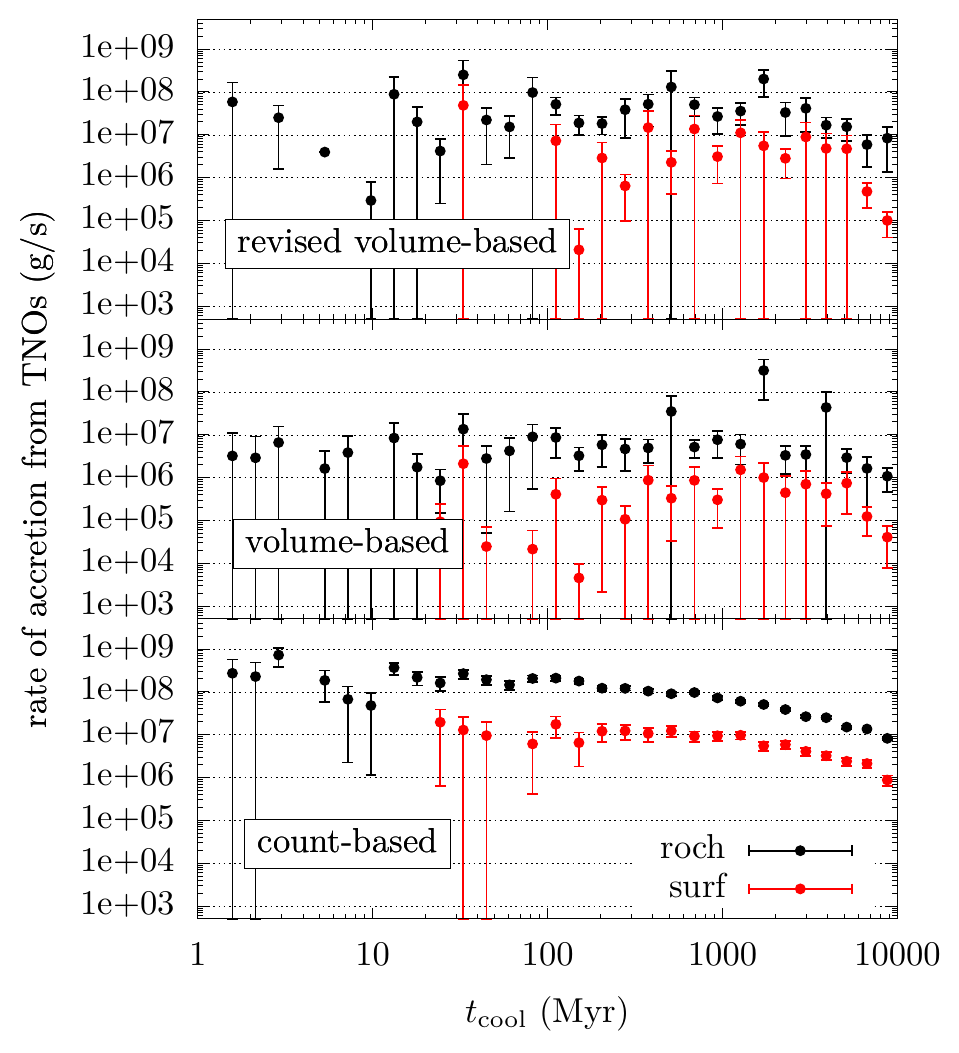}
\caption{Accretion rate of the solar WD from the TNOs as a function of WD cooling age. The symbol meaning and panel annotation are the same as Figure \ref{fig-mba-acc}. In the top panel, the accretion rate is called revised volume-based in that only contribution from the hot classicals, the resonant, the scattering, and the detached objects are included and the total mass of each subpopulation has been assigned specifically.}
\end{center}
\label{fig-tno-acc}
\end{figure}

The three types of accretion rates from the TNOs are presented in Figure \ref{fig-tno-acc}. Different from the MBA and JTA populations, we have nine runs for the TNOs and the mean rates of these runs are shown. The count-based accretion rate (bottom panel) shows a trend of slow and steady decrease. To be exact, the Roche accretion rate drops from a few times $10^8$ g/s for a cooling age of $\sim10$ Myr only by an order of magnitude to $\sim10^7$ g/s at 10 Gyr. The rate of surface accretion is an order of magnitude lower. The volume-based rate is, like those for MBAs and JTAs, overall lower than that of count-based and shows larger scatter reaching two orders of magnitude. Between a few hundreds of Myr and a few Gyr, the Roche rate remains around $10^7$ g/s while the surface rate oscillates around $10^5-10^6$ g/s. A clear decrease is only seen after that and at 10 Gyr, the two rates are $10^6$ g/s and $4\times10^4$ g/s, respectively. The revised volume-based accretion rate as shown in the top panel is in general higher by a factor of several compared to the volume-based rate. An obvious reason is that for the former, fewer objects (2150 compared 2844) share the same total mass so each is assigned a higher mass; but this is rather minor as the change in the number of bodies is only 25\%. We argue that the increase in accretion rate seen in the revised volume-based approach is physical as it better captures the biases for each subpopulation; for instance, the significant contribution from the scattered objects is better accounted in this scheme. Overall, the revised volume-based Roche accretion rate wanders around $10^7-10^8$ g/s from a few hundreds of Myr until a few Gyr while the surface rate is about $10^6-10^7$ g/s. Beyond, steady declines are seen and the two rates are $10^7$ g/s and $10^5$ g/s at 10 Gyr, respectively.

\subsubsection{Comparison with observations}
\begin{figure}
\begin{center}
\includegraphics[width=0.9\hsize]{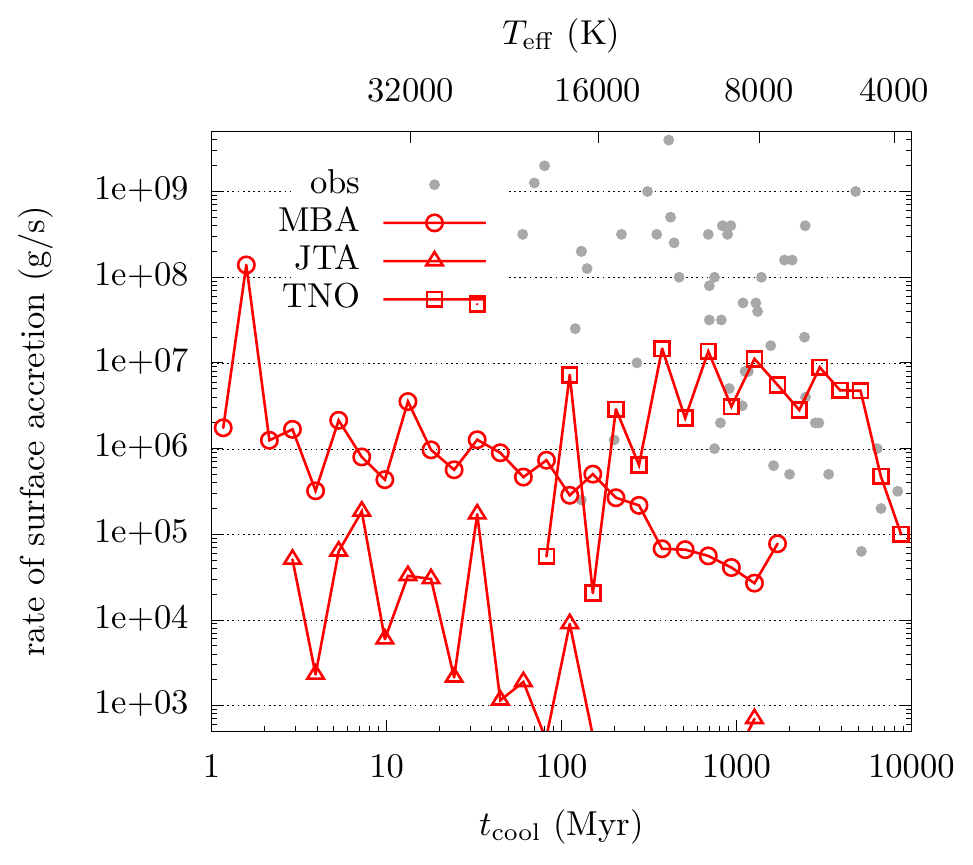}
\caption{Accretion rate of the solar WD from the MBAs (mass-based, unfilled circle), JTAs (volume-based, triangle), and TNOs (revised volume-based, square) as a function of WD cooling age. Here the rate is for surface accretion. The grey points are those for observed systems taken from \citet{Farihi2009} and references therein. The top horizontal axis shows the WD's effective temperature.}
\end{center}
\label{fig-wdacc-simp}
\end{figure}

Finally, the accretion rates from all the three small body populations are shown together in Figure \ref{fig-wdacc-simp} where we have used the mass-, volume-, and revised volume-based surface rates for the MBAs, JTAs, and TNOs, respectively. Before a cooling age of 100 Myr, MBAs (unfilled circle) dominate the accretion at a rate $\sim10^6$ g/s until 100 Myr. Beyond about 100 Myr, TNOs (square) take over and sustain a rate of $10^6-10^7$ g/s till several Gyr, after which it drops to $10^5$ g/s at 10 Gyr. JTAs (triangle) never play an important role.

Although extrasolar planetary systems are more diverse than simply all being Solar System clones, it is instructive to compare the accretion rates we obtain with observations. The observed rates from \citet{Farihi2009} and references therein are over-plotted in grey circles. Compared to the observations, the solar WD accretion we derive falls short by two orders of magnitude before a cooling age of 100 Myr. Later, while the rates for exoplanetary systems drop by an order of magnitude, the solar WD value actually increases by a factor of several at 100 Myr thanks to the TNOs. Therefore, from 1 Gyr and beyond, indeed the solar WD accretion rate agrees well with the extrasolar WDs and the decline in the accretion rate till 10 Gyr is well reproduced.

\section{Discussion and conclusion}\label{sec-dis}

We start with caveats to our study. The main sequence evolution of the solar system is omitted in this work. We first note that the formation of the three small body populations, the MBAs, JTAs, and TNOs have been greatly shaped by the early solar system history \citep[see][for a review]{Nesvorny2018} and since then they have been evolving under a quiescent environment for 4 Gyr. As a consequence, the dynamical \citep{Minton2010,DiSisto2014,Munoz-Gutierrez2019} and collisional \citep{Bottke2005,Hellmich2019,Durda2000} erosion over the next several Gyr before the Sun reaches the giant brach will be mild and the majority of mass contained (in large objects) within the three populations will not be affected.

During the giant brach (ML phase as called in this work), we have only considered the dynamical effects of the solar mass loss whereas the radiation force can be important as the central host may be (tens of) thousands of times more luminous than the present day Sun \citep{Sackmann1993}. Bodies $<$ 10 km within several au will be spun up to disruption \citep{Veras2014b,Veras2020c}. But the MBAs have a highly top heavy size frequency distribution and those $>$ 20 km account for more than 80\% of the total mass \citep[for example][]{Krasinsky2002}. Thus, the loss due to rotational fission during the giant branch is minor. At heliocentric distances beyond 30 au, only sub-km TNOs are influenced so this effect is negligible there.

The extreme solar luminosity also leads to the removal of volatile content within an object \citep{Jura2010,Malamud2016}. Within 10 au, objects of tens of km will become devoid of both free water and that embedded in rocks; large 100 km bodies may be able to retain a fraction of the embedded water. However, even for the relatively water-rich C-type asteroids, only $\lesssim10\%$ of their mass is contained in water \citep[see a recent discussion by][]{Beck2021}. Thus the loss of water will not affect the accretion rate significantly.

On the other hand, in our accretion rate estimate in Figure \ref{fig-wdacc-simp}, we have used the strict surface accretion criterion. When a small body reaches the Roche lobe of the WD, it will be tidally disrupted into a ring of fragments \citep{Debes2012,Veras2014,Li2021}. If it is a planet that scatters the small body close to the WD, the tidal fragments' orbits will probably intersect with that of the planet. As a result, a few tens of per cent of the fragments are scattered onto the WD in a few Myr by the planet \citep{Li2021}. Furthermore, if the parent body is large, $\gtrsim100$ km, sufficient pieces of tidal fragments are created and their mutual collision outpaces planet scattering. These tidal fragments will then be ground down to smaller and smaller collisional fragments, upon which Poynting-Robertson drag acts efficiently \citep{Li2021,Veras2015a}. As such, their orbits will be shrunk and circularised into a disk inside the Roche lobe, from which the WD accretes \citep{Rafikov2011,Rafikov2011a}. These effects are not modelled in this work but will likely cause additional metal accretion. Figure \ref{fig-acc-time} shows that tens of TNOs this large will reach the Roche lobe of the WD. Moreover, we have not considered the effect of the largest individuals on the other objects in the three small body populations. By gravitational scattering, these massive objects may cause a faster local diffusion \citep{Tiscareno2009,Munoz-Gutierrez2019} and therefore a (slightly) enhanced accretion rate onto the WD. In addition, we have omitted the Oort cloud objects which may contribute to WD pollution at a rate \citep{Veras2014a} perhaps comparable to the TNOs as derived here.

Another limitation of our model has to do with our simplistic implementation of the solar mass loss -- a linear decline with a timescale of 10, 20, or 40 Myr. In actuality, the mass loss may last for hundreds of Myr \citep{Sackmann1993}. This may eliminate the dominance by the MBAs in the accretion within a cooling age of about 100 Myr \citep[cf. also][]{Mustill2018} and therefore, TNOs may always dictate the solar WD accretion (also see Figure \ref{fig-acc-time} for the undetectability of accretion at young WD cooling ages). A much longer timescale for the solar mass loss also means a greater extent of loss of the three small body populations before the WD phase. As can be seen from Figure \ref{fig-gbloss}, even in a Gyr, the loss is only $\sim20\%$. Therefore, its effect on WD metal accretion is minor.

In summary, we believe that our estimate of the solar WD accretion rate remains quantitatively valid (perhaps within a factor of a few).

Debris disks are often proposed as the sources of WD metal pollution where the objects can be scattered toward the central host by planets \citep{Bonsor2011,Debes2012,Mustill2018,Smallwood2018,Smallwood2021,Veras2021}. Alternatively, several other authors worked on transporting asteroids toward the WD with the help of stellar companions, galactic tides and/or stellar flybys \citep{Veras2014a,Bonsor2015,Petrovich2017}. Most of those works, if deriving an absolute accretion rate, have adopted, in our terminology, a count-based approach \citep[perhaps except for][]{Debes2012}. Section \ref{sec-res-acc} shows that in a realistic small body population with object sizes covering a wide range \citep[see also][]{Wyatt2014}, the more physical mass (or volume as a good proxy)-based accretion rate is an order of magnitude smaller than the count-based rate. Therefore, the accretion rates obtained by those authors may need re-assessing and an order-of-magnitude correction may be necessary.

In unstable multi-planet systems \citep{Debes2002,Veras2013,Mustill2014,Veras2016a}, the surviving planets may scatter planetesimals inward to the WD \citep{Frewen2014,Mustill2018,Veras2021}. To allow instability to occur, the inter-planet spacing must be relatively close; however, this spacing is often too wide for many of the observed exoplanetary systems \citep{Fabrycky2014}. As such, more than 90\% of the two- or three-planet systems \citep{Maldonado2020,Maldonado2020a} and $\gtrsim50\%$ of systems with higher multiplicity \citep{Maldonado2021} are stable through the WD phase. Here our work represents a concrete example of a planetary system with stable planets and weakly unstable small objects where the latter are sporadically flung onto the WD, causing metal pollution \citep[cf., also][]{Bonsor2011,Debes2012,Smallwood2018,Smallwood2021}. Also, old debris disks typically have masses of the order $\sim0.01M_\oplus$ \citep{Wyatt2008}, reminiscent of the trans-Neptunian belt. Therefore, if the planets in these debris disk bearing systems are as efficient as the four giant planets in our solar system in scattering objects inwards, the long-term WD accretion rate can be readily reproduced.

Figure \ref{fig-wdacc-simp} clearly indicates that at younger cooling ages ($<$ 100-200 Myr), inner dry rocky material, i.e., the MBAs, dominates the accretion onto the solar WD while for older ages, volatile-rich substance, i.e., the TNOs, contributes more and for a much more extended period of time. Extrasolar debris disks often have two components \citep[][]{Kennedy2014}, a prominent example being the HR 8799 system \citep{Su2009}, much resembling the MBAs and the TNOs of the solar system. Such a two-component structure could have been carved by planets in between \citep{Lazzoni2018}. For those systems, one might expect that when the host star becomes a WD, metal pollution would follow a pattern like our solar system in an inside-out manner: accretion of dry material starts early but is soon eclipsed by volatiles after a few hundreds of Myr. This is however at odds with the observations. The exoplanetary WDs have been mostly consuming dry bulk Earth like material at cooling ages from a few times 100 Myr to a few Gyr \citep[e.g.,][]{Gansicke2012,Jura2014,Xu2019} and only in rare cases has volatile-rich pollutant been detected \citep[for instance,][]{Farihi2013}. Also, we have checked these systems accreting volatiles and do not find any sign of being universally old (if a cooling age is available in the discovery paper). Perhaps ways to reconcile the incompatibility would be that the architecture of our solar system is atypical or that devolatilisation and spin-up have been more effective than discussed above. 

Finally, we summarise the main findings of this work as follows. In order to study the dominant contributor to the pollution of the Solar WD as a function of time, we have numerically investigated the evolution of three main small body populations, the MBAs, JTAs, and TNOs, through the Sun's giant branch and WD phases under the perturbation of the four giant planets. We find that while 40\% of JTAs are lost during the giant branch as the Sun loses its mass, MBAs and TNOs are mostly stable. During the WD phase, objects from all three populations are leaking, some reaching the WD and causing metal pollution. The contribution from the JTAs is always negligible. Accretion from the MBAs dominates for a cooling age $\lesssim100$ Myr and the accretion rate is $10^6-10^7$ g/s, lower than the observed values by two orders of magnitude. Consumption of the TNOs only kicks in from $100$ Myr onward. From then and till 10 Gyr, the accretion rate for TNOs drops from $10^6-10^7$ g/s to $10^4-10^5$ g/s, in excellent agreement with those of the extrasolar WDs. We show that accretion rate from a small body population with an actual size frequency distribution is only a tenth of that from one comprised solely of equal-size objects, the reason being the dominance in the mass budget by largest individuals.

\acknowledgments
{\noindent \it Acknowledgment}\\
The authors thank the anonymous referee for the insightful feedback. D.L. and A.J.M. acknowledge financial support from Vetenskapsrådet (2017-04945). D.L. is grateful to support from the National Natural Science Foundation of China (12073019). Computations were carried out at the center for scientific and technical computing at Lund University (LUNARC).

\end{document}